\documentclass[12pt,preprint,longnamesfirst]{aastex}
\usepackage{emulateapj5}

\newcommand{\kms}{\hbox{ km\thinspace s$^{-1}$}}    %kms -1

\slugcomment{Scheduled for the October 2004 issue of The 
Astronomical Journal}

\shorttitle{Structure of NGC\,7662}
\shortauthors{Guerrero et al.}

\begin{document}

%% LaTeX will automatically break titles if they run longer than
%% one line. However, you may use \\ to force a line break if
%% you desire.

\title{Physical Structure of Planetary Nebulae. II. NGC 7662}

%% Use \author, \affil, and the \and command to format
%% author and affiliation information.
%% Note that \email has replaced the old \authoremail command
%% from AASTeX v4.0. You can use \email to mark an email address
%% anywhere in the paper, not just in the front matter.
%% As in the title, use \\ to force line breaks.

\author{Mart\'{\i}n A.\ Guerrero\altaffilmark{1,2}, 
        Elizabeth G.\ Jaxon\altaffilmark{2}, and 
        You-Hua Chu\altaffilmark{2,3}}
\affil{$^1$ Instituto de Astrof\'{\i}sica de Andaluc\'{\i}a, CSIC, 
            Granada 18008, Spain}
\affil{$^2$ Astronomy Department, University of Illinois at 
            Urbana-Champaign, Urbana, IL 61801}

\email{mar@iaa.es, ejaxon@uiuc.edu, chu@astro.uiuc.edu}

%% Notice that each of these authors has alternate affiliations, which
%% are identified by the \altaffilmark after each name. Specify alternate
%% affiliation information with \altaffiltext, with one command per each
%% affiliation.
\altaffiltext{3}{Visiting Astronomer, Kitt Peak National Observatory, 
    National Optical Astronomy Observatories, which is operated by the 
    Association of Universities for Research in Astronomy, Inc. (AURA) 
    under cooperative agreement with the National Science Foundation.}

%% Mark off your abstract in the ``abstract'' environment. In the 
%% manuscript style, abstract will output a Received/Accepted line 
%% after the title and affiliation information. No date will appear 
%% since the author does not have this information. The dates will 
%% be filled in by the editorial office after submission.

\begin{abstract}
We present a study of the structure and kinematics of the triple-shell 
planetary nebula NGC\,7662 based on long-slit echelle spectroscopic 
observations and $HST$ archival narrow-band images.  
The structure of NGC\,7662 main nebula consists of a central cavity 
surrounded by two concentric shells.  
The bright inner shell has a non-negligible thickness and the highest 
density in the nebula. 
The outer shell is filled with nebular material whose density profile 
shows a $\propto r^{-1}$ drop-off in the inner regions of the shell 
and flattens in the outermost regions.  
Both the inner and the outer shells can be described as prolate 
ellipsoids:  the inner shell is more elongated, while the outermost
layer of the outer shell expands faster.  
The physical structure of NGC\,7662 is qualitatively consistent with 
the predictions of hydrodynamical simulations based on the 
interacting-stellar-wind models for planetary nebulae with a 0.605 
M$_\odot$ central star.  Models with a higher AGB wind velocity 
are needed for detailed comparisons.
\end{abstract}

%% Keywords should appear after the \end{abstract} command. The 
%% uncommented example has been keyed in ApJ style. See the 
%% instructions to authors for the journal to which you are 
%% submitting your paper to determine what keyword punctuation 
%% is appropriate.

%% Authors who wish to have the most important objects in their paper
%% linked in the electronic edition to a data center may do so in the
%% subject header.  Objects should be in the appropriate "individual"
%% headers (e.g. quasars: individual, stars: individual, etc.) with the
%% additional provision that the total number of headers, including each
%% individual object, not exceed six.  The \objectname{} macro, and its
%% alias \object{}, is used to mark each object.  The macro takes the object
%% name as its primary argument.  This name will appear in the paper
%% and serve as the link's anchor in the electronic edition if the name
%% is recognized by the data centers.  The macro also takes an optional
%% argument in parentheses in cases where the data center identification
%% differs from what is to be printed in the paper.


%% From the front matter, we move on to the body of the paper.
%% In the first two sections, notice the use of the natbib \citep
%% and \citet commands to identify citations.  The citations are
%% tied to the reference list via symbolic KEYs. The KEY corresponds
%% to the KEY in the \bibitem in the reference list below. We have
%% chosen the first three characters of the first author's name plus
%% the last two numeral of the year of publication as our KEY for
%% each reference.

\section{Introduction}

Planetary nebulae (PNs) exhibit a wide variety of morphologies 
including round, elliptical, bipolar, and point-symmetric shapes 
\citep{B87,SCM92,M96}.
These main  morphological classes can be accounted for by the 
interacting-stellar-winds model \citep{KPF78} assuming that a 
pole-to-equator density gradient exists in the Asymptotic Giant 
Branch (AGB) wind \citep{B87}.  
The detailed morphology of some PNs has proven to be very complex 
and rich, however.  
By analyzing the morphology and kinematics of a PN, it is possible 
to establish its physical structure and gain insight into its 
evolutionary stage.  

We have started a program to analyze high-resolution images
and high-dispersion spectra of PNs in multiple nebular lines,
in order to determine the physical structure of PNs.
By comparing the results with hydrodynamical models, we hope to
understand better the formation and evolution of PNs.
The first PN we have studied is the Owl Nebula 
\citep[][hereafter Paper I]{Getal03}.
The second PN we have selected to study is NGC\,7662, which is
distinct from the Owl Nebula not only in morphology but also
in excitation structure.

NGC\,7662 has a bright main nebula that shows elliptical, 
double-shell morphology which is successfully reproduced by 
the interacting-stellar-winds model \citep[e.g.,][]{FM94}. 
The high surface brightness and moderately large size of this nebula, 
30\arcsec\ in diameter, make it ideal for a detailed case study of 
the physical structure of elliptical PNs.
We have used narrow-band images and high-dispersion spectra of
NGC\,7662 to build a 3-D spatio-kinematic model, derived the
density and ionization fraction profiles, and interpreted the
results in the light of the interacting-stellar-winds model.
This paper reports our study of NGC\,7662.

\section{Observations}

The {\it Hubble Space Telescope} ({\it HST}) archive contains images 
of NGC\,7662 obtained through narrow-band filters in the H$\alpha$, 
[O~{\sc iii}] $\lambda$5007, He~{\sc ii} $\lambda$4686, and 
[N~{\sc ii}] $\lambda$6584 lines (Program ID \#6117, PI: B.\ Balick; 
Program ID \#6943, PI: S.\ Casertano; Program ID \#8390, PI: A.\ Hajian). 
These filters have central wavelengths and FWHM bandwidths of 6565.6 \AA\ 
and 29 \AA\ (H$\alpha$), 5012.9 \AA\ and 37 \AA\ ([O~{\sc iii}]), 4694.5 
\AA\ and 35 \AA\ (He~{\sc ii}), and 6590.9 \AA\ and 40 \AA\ ([N~{\sc 
ii}]), respectively.  
We retrieved all archival \emph{HST} images of NGC\,7662 in these 
lines and combined them respectively to result in total exposure 
times of 200 s (H$\alpha$), 500 s ([O~{\sc iii}]), 200 s 
(He~{\sc ii}), and 2,800 s ([N~{\sc ii}]). 
The [N~{\sc ii}] image is seriously contaminated by the bright 
H$\alpha$ flux transmitted at the blue wing of the [N~{\sc ii}] 
filter because NGC\,7662 is a high-excitation PN with 
[N~{\sc ii}]/H$\alpha$ as low as $\sim$0.007 in the shells and 
0.116 in the [N~{\sc ii}]-bright knots \citep{HA97}.
We have thus used the transmission curve of the [N~{\sc ii}] 
filter to estimate this H$\alpha$ contribution and subtracted 
a scaled H$\alpha$ image from the [N~{\sc ii}] image to produce 
a net [N~{\sc ii}] image. 
The H$\alpha$, [O~{\sc iii}], He~{\sc ii}, and the net 
[N~{\sc ii}] images are presented in Figure~1.   

High-dispersion spectroscopic observations of NGC\,7662 were obtained 
with the echelle spectrograph on the 4-m telescope at the Kitt Peak 
National Observatory (KPNO) on 1986 October 21--23.  
The spectrograph was used in the single-order, long-slit mode, covering 
only the [O III] $\lambda$4959, 5007 lines.  
The 79 line mm$^{-1}$ echelle grating and the long-focus red camera 
were used; the resultant reciprocal dispersion was 2.47 \AA~mm$^{-1}$.  
The spectra were recorded with the 800$\times$800 TI $\# 2$ CCD with a 
2$\times$2 binning.  
The resulting pixel size of 30 $\mu$m corresponded to a spatial scale 
of 0\farcs33~pixel$^{-1}$ and a sampling of 4.5 km s$^{-1}$ 
pixel$^{-1}$ along the dispersion direction.  
A slit width of 150 $\mu$m (1\arcsec) was used, and the resultant 
instrumental FWHM was 9.5$\pm$0.3 km s$^{-1}$.  
The angular resolution, determined by the seeing, was 1\farcs9. 
 
We obtained four long-slit spectra of NGC\,7662: 
one along the major axis (PA=44\arcdeg), two parallel to but with 
a 3\arcsec\ offset from the major axis, and one along the minor 
axis (PA=135\arcdeg).  
The slit positions are marked on the [O~{\sc iii}] image
in Figure~1. 
The exposure times were 300 s for the spectrum along the major axis, 
and 200 s for the other three.  
The spectra were reduced using standard IRAF packages\footnote{
IRAF is distributed by the National Optical Astronomy 
Observatories, which is operated by the Association 
of Universities for Research in Astronomy, Inc.\ (AURA) 
under cooperative agreement with the National Science 
Foundation.}.
The curvature-corrected, wavelength-calibrated echellograms are
displayed in Figure 2.  
The velocities are referenced to the systemic velocity of 
NGC\,7662.

\section{Results}

\subsection{Morphology}

NGC\,7662 is a triple-shell nebula with a bright 
17\farcs9$\times$12\farcs4 inner shell enveloped 
by a 30\farcs8$\times$27\farcs2 outer shell and 
surrounded by a faint spherical halo 134\arcsec\ 
in diameter \citep{C87}.
The major-to-minor axial ratio of the inner shell, $\sim$1.4,
is noticeably larger than that of the outer shell, $\sim$1.1.
The inner shell appears as a bright ``rim''\footnote{The inner 
shell is called ``rim'' by \citet{FBR90}} with an apparent 
thickness of $\sim$2\farcs5, about half the length of the 
semi-minor axis.
In contrast, the outer shell appears as an envelope attached 
to the inner shell, showing no limb-brightening.

The morphology of the main nebula is complicated by the 
presence of ``fast low-ionization emission regions (FLIERs)," 
as discussed in detail by \citet{B98}.
Our aim is to study the basic underlying structure of the nebula, 
therefore we select the sectors of the main nebula that are free 
of FLIERs.  
In particular, the analysis of the density and ionization structure 
of the nebula in Section $\S$3.3 will be focused in its northwest 
quadrant.

\subsection{Kinematics}

The kinematics of NGC\,7662 has been studied by \citet{BPI87}
using echelle spectra of the H$\alpha$ and [\ion{N}{2}] lines.  
As the H$\alpha$ line has a large thermal width and the 
[\ion{N}{2}] emission is dominated by the FLIERs, we have 
chosen to use [O~{\sc iii}] to study the kinematics.

Our [O~{\sc iii}] echellograms, displayed in Figure 2, show 
two kinematic systems corresponding to the inner and outer 
shells.  
The inner shell appears as a hollow position-velocity ellipse 
that has a marked tilt along the major axis (PA $45\arcdeg$), 
indicating that it is an expanding prolate ellipsoidal shell 
with its northeast side tilted towards us.  
The velocity dispersion within the shell is small; 
for example, the blue- and red-shifted components at the central 
position have an observed velocity FWHM of 13 km~s$^{-1}$, 
corresponding to an intrinsic velocity FWHM of 8 km~s$^{-1}$ 
(after subtracting the instrumental and thermal widths).  
The brightness variations along the position-velocity ellipse 
can be naturally produced by the different line-of-sight path 
lengths through a prolate ellipsoidal shell \citep{FM94}.

The outer shell also appears as a position-velocity ellipse, 
with a hollow center and thick filled rims, but without an 
obvious tilt.  
This echelle line morphology is consistent with a thick 
expanding shell with a small ellipticity.  
For sightlines through the thick shell rims, the broad line 
profiles are caused by the wide range of projection angles 
between the radial expansion velocity and the line of sight.  
Near the shell center, the line profiles of the outer shell 
are superimposed on those of the bright inner shell;  
however, the narrow widths of the blue- and red-shifted 
components indicate that the outer shell expands at least 
as fast as the inner shell and that the velocity dispersion 
in the outer shell is small as well.  
The brightness variations within the thick shell rims in the 
echellograms can be caused by a combination of small density 
and velocity variations within the shell.  
The echellograms also show faint unresolved emission beyond the 
visible boundary of the outer shell.  
This extension originates from the faint halo of NGC\,7662,
and its FWHM $\sim$ 45 \kms\ is unusually large for faint
PN halos \citep{CJ89,GVM98}.

The observed position-velocity ellipse of an expanding shell can 
constrain its geometry and kinematics.  
Assuming a simple homologously expanding ellipsoidal shell 
(i.e., $v_{\rm exp} \propto r$), it is possible to derive 
the axial ratios, inclination angles and expansion velocities
of the shell.  
For the inner shell of NGC\,7662, the observed position-velocity 
ellipse is narrow and well defined, thus it can be directly 
compared with synthetic position-velocity ellipses to determine 
its geometry and kinematics.  
The outer shell is more difficult to model because its thick shell 
can be divided into multiple thin expanding shells, but the 
position-velocity ellipses of these thin shells cannot be resolved 
from one another.  
In the Owl Nebula, it has been shown that the outermost layer of 
its outer shell (most well defined in the low ionization 
[N~{\sc ii}] line) is characterized by a position-velocity ellipse 
that encloses the echelle line image of that shell (Paper I).  
We have thus adopted such an ellipse as the position-velocity 
ellipse of the outermost layer of the outer shell of NGC\,7662.
Note that this assumption would be valid as long as there is 
no negative velocity gradient in the outer shell

Using these models, we have produced synthetic position-velocity 
ellipses to match these observed in the inner shell and the
outermost layer of the outer shell.  
The position-velocity ellipses of the best-fit models are 
overplotted in Figure~2, and the relevant parameters are 
summarized in Table~1.  
For the inner shell, we find that its major axis is tilted by 
50\arcdeg\ against the line of sight along PA 45\arcdeg, and 
that its expansion velocities are 35$\pm$3 km~s$^{-1}$ and 
53$\pm$6 km~s$^{-1}$ in the equatorial plane and along the 
pole, respectively.  
For the outer shell, the small line tilt implies a small ellipticity 
or a pole-on geometry.  
To overcome this ambiguity, we assumed that the inclination of the 
symmetry axis of the outer shell is the same as that of the inner 
shell \citep{B98} and derived equatorial and polar expansion 
velocities for the outer shell of 50$\pm$5 km~s$^{-1}$ and 
60$\pm$8 km~s$^{-1}$, respectively. 
The outer shell of NGC\,7662 has a smaller axial ratio than 
the inner shell, 1.2 versus 1.5.  
The outermost layer of the outer shell expands faster than 
the inner shell, which may be expected for dense gas diffusing 
into the surroundings with much lower density.  
The inner shell has a kinematic age of $\sim$700$\times\frac{d}{800}$ 
yr, where $d$ is the distance in pc \citep[$d$=800 pc,][]{HT96}, and 
the outer shell has a kinematic age of $\sim$1,050$\times\frac{d}{800}$ 
yr.

\subsection{Density and Ionization Structure}

The spatio-kinematic models of NGC\,7662 deduced in $\S$3.2 can be 
used along with the observed surface brightness distribution in 
different emission lines to derive the density and ionization 
structure within this nebula.  
The northwest quadrant of NGC\,7662 provides the most appropriate 
region for this analysis because it has no FLIERs and shows smooth, 
featureless emission in both the direct images and [O~{\sc iii}] 
echellogram.  
Furthermore, this quadrant contains the minor axis along which the 
surface brightness distribution is less sensitive to the detailed 
shape and orientation of the nebular shells.  
The observed surface brightness profiles of the H$\alpha$, 
[O~{\sc iii}], He~{\sc ii}, and [N~{\sc ii}] lines along 
this minor axis are displayed in Figure~3.  
Below is our analysis of the shell structure of NGC\,7662 
along the minor axis in the northwest quadrant.

To derive the density and ionization structure from the observed 
surface brightness, we have adopted a distance of 800 pc \citep{HT96},
a logarithmic extinction $c{\rm (H}\beta${\rm )}=0.10, and an 
electron temperature of 12,500 K \citep{HA97}.  
We have approximated the main nebula of NGC\,7662 by two ellipsoidal 
shells with the sizes, axial ratios and inclinations given in Table~1.  
We have further assumed that each shell consists of concentric 
sub-shells of constant density and ionization.  
As the surface brightness at a given radius $r$ includes 
contributions from sub-shells of radii $\ge r$, the 
derivation of the density is best done from outside in.

The surface brightness profile of the H$\alpha$ line is used to 
determine the radial profile of density $n(r)$.  
We adopt a power law for the density profile, i.e., 
$n(r) = n(r_0) r^{-i}$, where $n(r_0)$ and $i$ will be 
determined from the fit to the surface brightness profile.  
The presence of multiple kinks in the H$\alpha$ surface brightness 
profile implies that the density profile cannot be described by a 
single power law throughout the nebula.  
We have thus derived the density profile piecewisely.  
We experimented various power-law index $i$ and find that $i=0$ 
and 1 provide the best fits for the inner and outer shell, 
respectively.  
The best-fit density profile is shown in Figure~4.  
The corresponding surface brightness profile, overplotted in 
Figure~3, is in good agreement with the observed H$\alpha$ 
profile from radii of 2\arcsec\ to 14\arcsec.

It is interesting to note that the density profile in the outer 
shell is better described by a power law index of 1, instead of 
2, as expected for an AGB wind with constant velocity and 
mass-loss rate.
Three segments marked by small density jumps at radii $\sim$10$\farcs8$ 
and $\sim$12$\farcs2$ are present in the density profile of the outer 
shell.  
From the outer shell to the inner shell there is a sharp increase in 
density.  

The inner shell is best described with a constant density $\sim$5,000 
cm$^{-3}$ over a thickness $\gtrsim 2\arcsec$, but the density drops 
off steeply at its inner edge.  
This shell density is within the range determined from 
spectrophotometric observations of the [S~{\sc ii}] doublet, 
$\sim$4,000 cm$^{-3}$ for the inner shell and $>$6,000 cm$^{-3}$ 
in isolated dense regions \citep{LP96,HA97}.  
The density in the central cavity is low, but it cannot be 
determined with certainty because the surface brightness at 
the central region of the nebula is overwhelmed by the emission 
from the inner and outer shells.  
The sharp leading edge of the inner shell is in agreement with its 
supersonic expansion within the inner edge of the outer shell.  

The density profile and the assumed ellipsoidal shell geometry can be 
used to determine the ionized mass in the inner and outer shells.  
The mass in the inner shell is $\sim$0.01 M$_\odot$, while that of the 
outer shell is $\sim$0.04 M$_\odot$. 
To test whether the inner shell consists entirely of swept-up AGB wind, 
we have extrapolated the density profile of the outer shell to the 
center and determined the swept-up mass.  
An excellent agreement exists between the expected swept-up mass 
and the mass in the inner shell, suggesting that no extraneous 
mass-loss is needed to form the inner shell.

The ionization structure of NGC\,7662 can be studied using the 
observed surface brightness profiles of the He~{\sc ii}, [O~{\sc iii}], 
and [N~{\sc ii}] lines (Fig.~3).  
The surface brightness, in conjunction with electron temperature and 
density, can be used to derive the ionic abundance.  
We have adopted appropriate electron temperatures for lines of 
different excitations: 
14,000 K and 11,000 K for the high-excitation [O~{\sc iii}] and 
He~{\sc ii} lines in the inner and outer shells, respectively, and 
8,500 K for the low-ionization [N~{\sc ii}] line \citep{B86,HA97}.
The ionic abundances determined using these temperatures and the above 
derived densities are then divided by the corresponding elemental 
abundances \citep{B86} to compute the ionization fractions of He$^{++}$, 
O$^{++}$, and N$^{+}$ shown in Figure~4.  

An inspection of Fig.~4 shows that the ionization fraction of He$^{++}$ 
is rather constant and greater than 0.8 through the nebula, while those 
of O$^{++}$ and N$^+$ increase outwards, as expected from stratified 
ionization.  
It must be noted that the detailed level and shape of the ionization 
fraction profiles depends on the choice of elemental abundances and 
electron temperatures.  
The large ionization fraction of He$^{++}$, for instance, may imply 
full ionization of He, as typically predicted by ionization models 
\citep[e.g.,][]{AB97}, with He abundances smaller by $\sim$10\% than 
those here adopted.   
Similarly, variations of the electron temperature with radius in 
NGC\,7662, as reported by \citet{B86}, may affect the derived 
ionization fraction profiles, especially those of O$^{++}$ and N$^+$, 
so that larger values of electron temperature mimic larger ionization 
fractions of O$^{++}$ and N$^+$.
The increased ionization fraction of O$^{++}$ in the outermost regions 
of NGC\,7662 seems real, however, as the electron temperature is known 
to decrease outwards \citep{B86}.  
Therefore, the sharp increase of the ionization fraction of O$^{++}$ 
at $r$ $\sim$ 7\arcsec\ signifies the transition of ionization stage 
at the edge of the O$^{+3}$ Str\"omgren sphere, although the exact 
shape of the O$^{++}$ ionization fraction profile and the radius of 
the O$^{+3}$ Str\"omgren sphere might be questionable.  

% 
% The ionization fraction profiles vary smoothly across the density 
% jump from the inner shell to the outer shell, but the O$^{++}$ and 
% He$^{++}$ ionization fraction show large jumps within the outer
% shell where the density variations are mild.
% The ionization potentials of O$^{++}$ and He$^{+}$ are 54.9 and 
% 54.4 eV, respectively.
% The sharp increase of the ionization fraction of O$^{++}$ at 
% $r$ $\sim$ 8\farcs5 probably signifies a transition of 
% ionization stage at the edge of the O$^{+3}$ Str\"omgren sphere.
% The rapid decline of the ionization fraction of He$^{++}$
% at $r$ = 11\arcsec\ to 12\farcs5 most likely signifies the
% edge of the He$^{++}$ Str\"omgren sphere.
% 

\section{Discussion}

Hydrodynamical simulations of the physical structure of PNs have been 
carried out in the framework of interacting stellar winds taking into 
account the evolution and mass-loss history of the central star 
\citep[e.g.,][]{FBR90,M94,VMG02,P04}.  
In these simulations, the fast stellar wind excavates a central cavity 
and snowplows the AGB wind to produce a double-shell nebula.  
The inner shell, being the swept-up AGB wind, is denser than the outer 
shell.  
The outer shell has a relatively flat density profile as its sudden 
ionization has resulted in strong pressure gradients that have modified 
the original $r^{-2}$ law. 
This overall structure is observed in NGC\,7662; 
the H$\alpha$ image and echelle spectra are similar to those 
simulated by \citet{M94}, and the mass in the inner shell is in 
agreement with the expected swept-up mass of the AGB wind.

We attempt to quantitatively compare the density profile and 
kinematic structure of NGC\,7662 to PN models.
The density profile of  NGC\,7662 is characterized by: \begin{itemize}
\itemsep-0.1cm
\item a density contrast of $\sim$2 between the outer and inner shells,
\item an inner shell with a thickness of 0.012 pc, about 0.3--0.5 times
     the shell radius, and
\item an almost flat density profile in the outer shell, significantly
    flatter than the $r^{-2}$ law.
\end{itemize}
The radial profile of the expansion velocity of NGC\,7662 cannot be 
determined observationally because of the limited spatial resolution 
of the echelle observations and the non-negligible thermal line widths.
Our analysis in $\S3.2$ provides only the following basic 
kinematic characteristics of NGC\,7662: \begin{itemize}
\itemsep-0.1cm
\item the inner shell has a small internal velocity dispersion, and
    expands at 35\kms\ and 53\kms\ along the equator and poles,
    respectively; and 
\item the outer shell also has small internal velocity dispersions, 
    and expands at similar velocities to the inner shell, but its 
    outermost layer expands faster, at 50\kms\ and 60\kms\ along the 
    equator and poles, respectively.  
\end{itemize}

We have compared the density profile and kinematic structure of 
NGC\,7662 to those synthesized from the \citet{P04} models of 
spherical PNs that take into account different initial conditions 
and central star masses.
They have divided the PN evolution into four phases: 
Phase I corresponds to the youngest stage when the central star is still
too cool to ionize the AGB wind;  
Phase II corresponds to the compression stage when an ionized double-shell 
structure is visible; 
Phase III is the stage when the central star reaches the maximum 
temperature and the inner-to-outer shell density contrast is the highest; 
and in Phase IV recombination starts and a single shell is observed 
possibly with a recombination halo.
The extended outer shell, high densities, density profile,
and velocity structure of NGC\,7662 are compatible with those of 
a PN at Phase II evolving toward Phase III (compare to Fig.~7
of Perinotto et al.).

\citet{P04} modeled PNs for different central-star masses and
presented their results in Figures 13--16.  
We can rule out a central star mass much below 0.6 M$_\odot$ 
because the densities in NGC\,7662, a few 10$^3$ cm$^{-3}$, 
are more than an order of magnitude higher than those expected 
from PNs with low-mass central stars.
We can also rule out a central star mass as high as 0.8--0.9 
M$_\odot$, because such PNs have higher shell densities and
shorter double-shell lifetime than NGC\,7662.
The observed properties of NGC\,7662 are most compatible 
with the model for a 0.605 M$_\odot$ central star.

We can also compare NGC\,7662 to the \citet{P04} models for
a 0.605 M$_\odot$ central star with different AGB mass loss rates
(their Fig.~10).  
The density and velocity profiles of NGC\,7662 are most compatible
with the model with an AGB mass loss rate of $3\times10^{-5}$
M$_\odot$ yr${-1}$ and wind velocity of 10 \kms; however, the 
expansion velocities of NGC\,7662 are much higher than the 
$\sim$15 km s$^{-1}$ expansion velocities from this model.
A higher shell expansion velocity can be achieved if the AGB wind
velocity is higher, and to maintain the same AGB wind density profile
a proportionally higher AGB mass loss rate is required.
It is possible that the AGB wind velocity of NGC\,7662 was 30--50\kms,
and the AGB mass loss rate was $\sim10^{-4}$ M$_\odot$ yr$^{-1}$.
Future models of NGC\,7662 ought to consider these AGB wind properties.

\section{Summary}

NGC\,7662 is a triple-shell PN with a bright inner shell (or rim)
surrounded by an outer shell and a faint, large halo.
We have analyzed high-resolution images and spectra of the inner
and outer shells of NGC\,7662, and constructed 3-D spatio-kinematic
models for its inner shell and the outermost layer of the outer shell.
Both can be modeled as expanding prolate ellipsoidal shells with the
northeast end of the major axis tilted toward us.

We have further used the spatio-kinematic models and the H$\alpha$,
\ion{He}{2}, [\ion{O}{3}], and [\ion{N}{2}] surface brightness profiles
to determine the density profile and ionization structure along the 
minor axis.
The density and kinematic structure of NGC\,7662 are compared to
\citet{P04} models of spherical PNs.
We find that NGC\,7662 is in their Phase II evolutionary stage (the
compression stage), as its outer shell is still fairly extended, but
may be close to Phase III evolutionary stage as its inner shell
does not seem to be compressed strongly from its interior.
NGC\,7662 is most compatible with models for a central star mass
of 0.605 M$_\odot$.
Future models with higher AGB wind velocities, such as 30--50\kms,
are needed for detailed comparisons with NGC\,7662.

\acknowledgments

M.A.G.\ acknowledges support from the grant AYA~2002-00376 of
the Spanish MCyT (cofunded by FEDER funds).
Some of the data presented in this paper were obtained from the 
Multimission Archive at the Space Telescope Science Institute (MAST).  
STScI is operated by the Association of Universities for Research in 
Astronomy, Inc., under NASA contract NAS5-26555.  

%% To help institutions obtain information on the effectiveness of their
%% telescopes, the AAS Journals has created a group of keywords for telescope
%% facilities. A common set of keywords will make these types of searches
%% significantly easier and more accurate. In addition, they will also be
%% useful in linking papers together which utilize the same telescopes
%% within the framework of the National Virtual Observatory.
%% See the AASTeX Web site at http://www.journals.uchicago.edu/AAS/AASTeX
%% for information on obtaining the facility keywords.

%% After the acknowledgments section, use the following syntax and the
%% \facility{} macro to list the keywords of facilities used in the research
%% for the paper.  Each keyword will be checked against the master list during
%% copy editing.  Individual instruments can be provided in parentheses,
%% after the keyword, but they will not be verified.

% Facilities: \facility{KPNO(4-m echelle)}, \facility{HST(WFPC)}.

\begin{deluxetable}{lcccccr}
\tablenum{1}
\tablewidth{36.5pc}
\tablecaption{Spatio-Kinematic Model of the Inner and Outer Shells 
of NGC\,7662}
\tablehead{
\multicolumn{1}{c}{} & 
\multicolumn{1}{c}{$V_{\rm equatorial}$} & 
\multicolumn{1}{c}{$V_{\rm polar}$} & 
\multicolumn{1}{c}{Inclination\tablenotemark{\dagger}} & 
\multicolumn{1}{c}{Angular Size} & 
\multicolumn{1}{c}{Linear Size\tablenotemark{\ddagger}} & 
\multicolumn{1}{c}{Kinematic Age\tablenotemark{\ddagger}} \\ 
\multicolumn{1}{c}{} & 
\multicolumn{1}{c}{[km~s$^{-1}$]} & 
\multicolumn{1}{c}{[km~s$^{-1}$]} & 
\multicolumn{1}{c}{} & 
\multicolumn{1}{c}{} & 
\multicolumn{1}{c}{[pc]} & 
\multicolumn{1}{c}{[yr]}}
\startdata 
Inner Shell & 35 & 53 & ~50\arcdeg & 13\arcsec$\times$19\farcs7 & 0.050$\times$0.076 &   700~~~~~~ \\
Outer Shell & 50 & 60 & ~50\arcdeg & 28\arcsec$\times$33\farcs6 & 0.109$\times$0.130 & 1,050~~~~~~ \\
\enddata
\tablenotetext{\dagger}{Angle between the major axis and the line of sight.}
\tablenotetext{\ddagger}{Adopting a distance to NGC\,7662 of 800 pc.}
\end{deluxetable}

\clearpage

%% No more than seven \figcaption commands are allowed per page,
%% so if you have more than seven captions, insert a \clearpage
%% after every seventh one.

%% There must be a \figcaption command for each legend. Key the text of the
%% legend and the optional \label in curly braces. If you wish, you may
%% include the name of the corresponding figure file in square brackets.
%% The label is for identification purposes only. It will not insert the
%% figures themselves into the document.
%% If you want to include your art in the paper, use \plotone.
%% Refer to the on-line documentation for details.

\begin{figure}
\epsscale{1.0}
\plotone{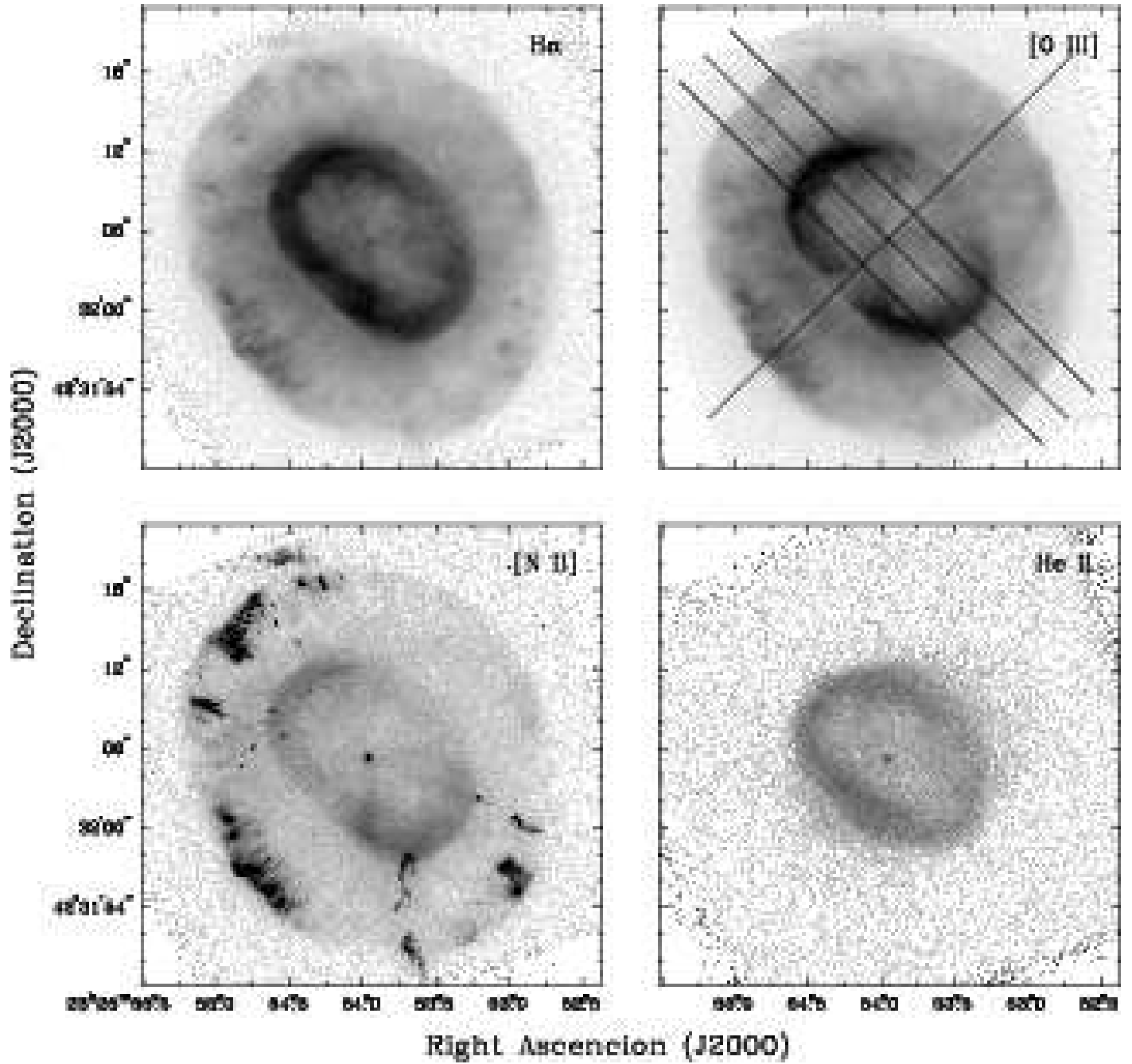}
\caption{
Negative grey-scale representation of the $HST$ WFPC2 narrow-band images 
of NGC\,7662.  
The contribution of the H$\alpha$ line in the [N~{\sc ii}] $\lambda$6584 
filter has been subtracted.  
The slit positions of the echelle observations are marked on the 
[O~{\sc iii}] image.  
The image displays are square root.  
\label{images}}
\end{figure}

\begin{figure}
\epsscale{0.9}
% edit 2spec.ps from ``90 rotate'' to ``0 rotate''
% ps2epsi 2spec.ps 2spec.eps
% edit bounding box to
%%BoundingBox: 390 31 925 350
\plotone{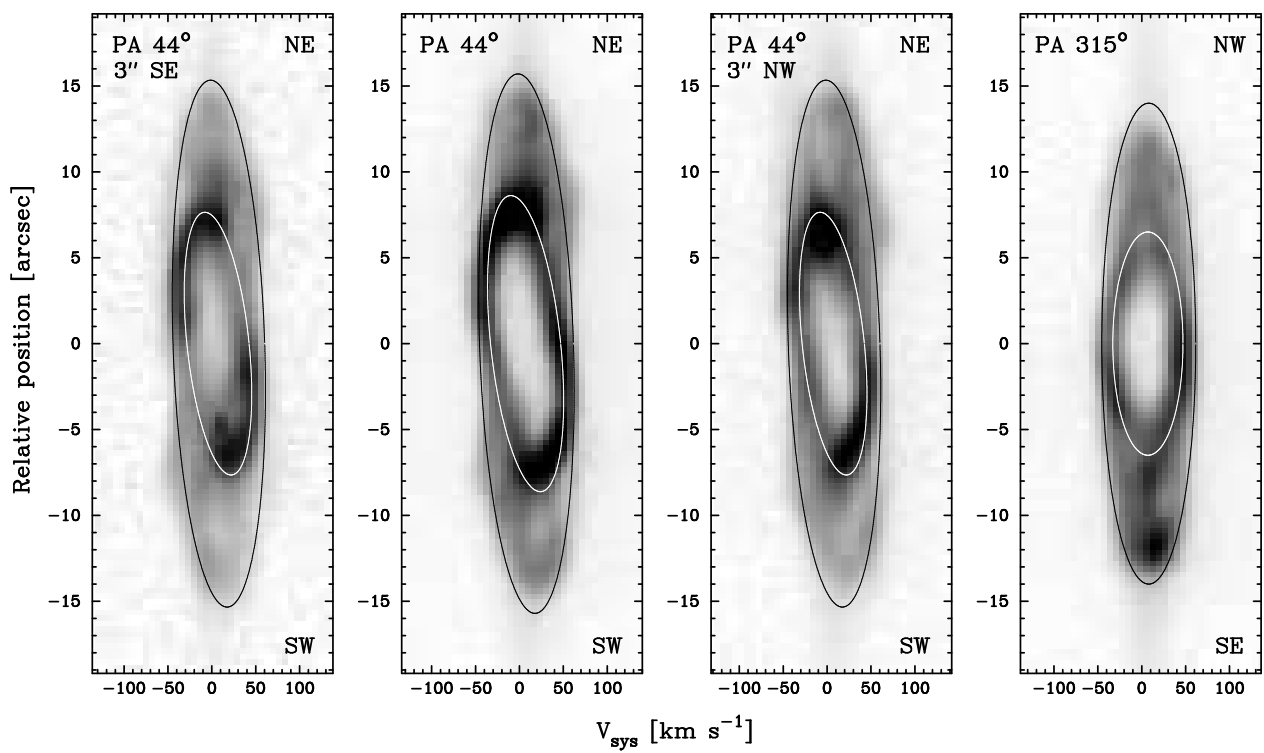}
\caption{
Negative grey-scale representation of the [O~{\sc iii}] $\lambda$5007 
emission line echellograms at PAs 44\arcdeg\ and 315\arcdeg.  
The ellipses overlaid on the echellograms correspond to the best-fit 
models for the inner and outer shells described in $\S3.2$ (displayed 
in white and black, respectively).  
\label{spec}}
\end{figure}

\clearpage

\begin{figure}
\epsscale{0.6}
\plotone{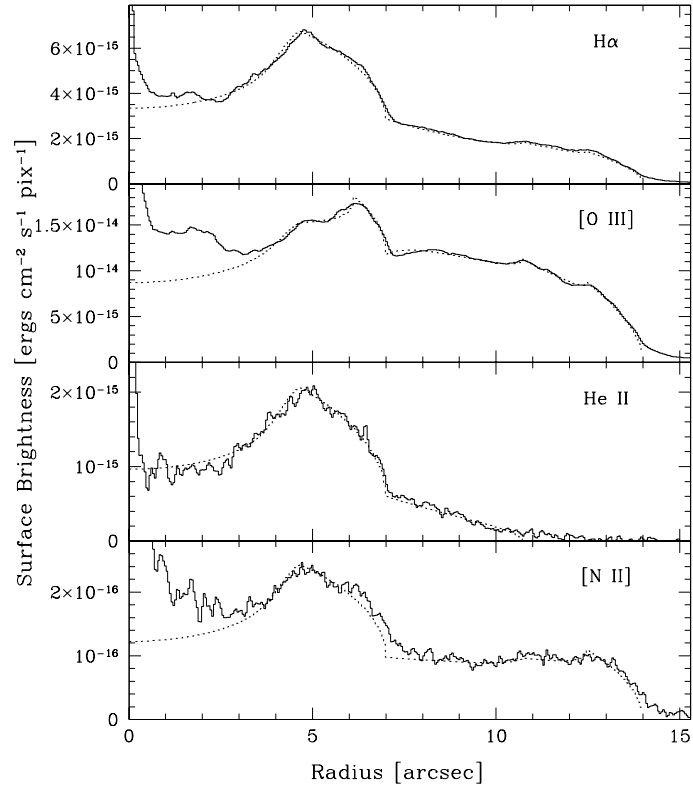}
\caption{
Surface brightness profiles along PA~315\arcdeg\ of the H$\alpha$, 
[O~{\sc iii}] $\lambda$5007, He~{\sc ii} $\lambda$4686, and [N~{\sc ii}] 
$\lambda$6584 lines (solid curves) extracted from the $HST$ narrow-band 
images of NGC\,7662.  
The dotted curves are our synthetic surface brightness profiles of 
these lines.  
\label{profile}}
\end{figure}

\begin{figure}
\epsscale{0.5}
\plotone{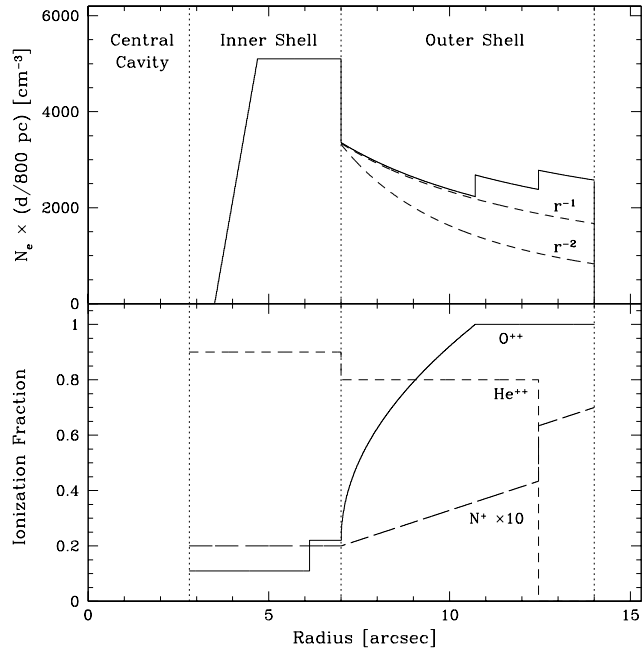}
\caption{
Radial profiles of the best-fit density and O$^{++}$, He$^{++}$ and 
N$^+$ ionization fractions of NGC\,7662 along PA~315\arcdeg.  
The positions of the central cavity, inner shell and outer shell 
are marked by vertical dotted lines.  
The dashed lines in the plot of the density profile correspond to a 
density dependence $\propto r^{-1}$ and $\propto r^{-2}$ for the outer 
shell.  
In the plot of the ionization fraction profiles, the N$^+$ ionization 
fraction has been multiplied by 10 and plotted in long dashes.  The
O$^{++}$ and He$^{++}$ ionization fractions are plotted in solid
and short-dahsed curves, respectively.
\label{nx}}
\end{figure}

\end{document}